\documentclass[aps,twocolumn,prl,showpacs,preprintnumbers]{revtex4}
\usepackage{graphicx}
\usepackage{color}

\begin{document}

\title{Combined effect of doping and temperature on the anisotropy of low-energy plasmons in monolayer graphene}

\author{ Godfrey Gumbs,$^{1,2}$ Antonios Balassis,$^3$ and V. M. Silkin$^{1,4,5}$}

\affiliation{
$^1$Donostia International Physics Center (DIPC), Paseo de Manuel Lardizabal 4, 20018 San Sebasti\'an/Donostia, Spain\\
$^2$ Department of Physics and Astronomy, Hunter College of the City
University of New York, 695 Park Avenue, New York, NY 10065, USA\\
$^3$Department of Physics and Engineering Physics, Fordham University, 441 East Fordham Road, Bronx, New York 10458, USA\\
$^4$Departamento de F\'{\i}sica de Materiales, Facultad de Ciencias
Qu\'{\i}micas, Universidad
del Pa\'{\i}s Vasco, Apdo. 1072, 20080 San Sebasti\'an/Donostia, Spain\\
$^5$IKERBASQUE, Basque Foundation for Science, 48011, Bilbao, Spain
}

\date{\today}

\begin{abstract}
We compare the two-dimensional (2D) plasmon dispersion relations for
monolayer graphene when the sample is doped with carriers in the
conduction band and the temperature $T$ is zero with the case when
the temperature is finite and there is no doping. Additionally, we
have obtained the plasmon excitations when there is doping at finite
temperature. The results were obtained in the random-phase
approximation which employs energy electronic bands calculated using
{\it ab initio} density functional theory. We found that in the
undoped case the finite temperature results in appearance in the
low-energy region of a 2D plasmon which
is absent for the $T=0$ case. Its energy is gradually increased
with increasing $T$. It is accompanied by expansion in the momentum
range where this mode is observed as well. The 2D plasmon dispersion
in the $\Gamma$M direction may differ in substantial ways from that
along the $\Gamma$K direction at sufficiently high temperature and
doping concentrations. Moreover, at temperatures exceeding
$\approx300$ meV a second mode emerges along the $\Gamma$K direction
at lower energies like it occurs at a doping level exceeding
$\approx 300$ meV. Once the temperature exceeds $\approx 0.75$ eV
this mode ceases to exit whereas the 2D plasmon exists as a
well-defined collective excitation
up to $T=1.5$ eV, a maximal temperature investigated in this work.

\end{abstract}


\maketitle

\section{Introduction}

Graphene,   consisting of a single layer of carbon atoms, is an
ideal realization of a system in which electrons, confined in two
dimensions, are quantum mechanically enhanced
\cite{drprb74,najpsj76,chmiprb91,chgoprb91,chmiprb92}. Moreover,
recent advances in fabrication and micromechanical extraction
techniques for graphite structures now make it possible for such
exotic two-dimensional (2D) electron systems to be probed
experimentally. The collective quasiparticle phenomena giving rise
to the plasmon excitation spectra may display interesting features
which are accessible experimentally
\cite{bupssb77,krhaprl08,luloprb09,kreiprb10}. Additionally, their
behavior is expected to differ substantially from  the
well-understood plasmonic properties for quantum wells in
conventional semiconductor heterostructures, including group-IV
compounds, binary systems of group III-IV elements, metal
chalcogenides and complex oxides \cite{buhoan13,xulicr13}. This
difference is due to the unique electronic properties of graphene
which possesses electron-hole (e-h) degeneracy and zero carrier
effective mass near the K point at the corner of the Brillouin zone
(BZ) \cite{wapr47}. Indeed, at zero temperature $T$, a low-frequency
2D plasmon mode with energy dispersion $\omega_{2D}\propto q^{1/2}$
has been obtained \cite{wustnjp06,hwdaprb07} for doped graphene with
the use of an isotropic Dirac cone approximation (DCA).
We note that the electronic states
of graphene at the Dirac point can be described within the framework
of basic numerical schemes. Graphene consists of a flat layer of
carbon atoms arranged in a hexagonal lattice with two carbon atoms
per unit cell. Of the four valence states, three $sp^2$ orbitals
form a $\sigma$ state with three neighboring carbon atoms. One $p$
orbital emerges as delocalized $\pi$ and $\pi^*$ states which
constitute the highest occupied valence and the lowest unoccupied
conduction bands. The $\pi$ and $\pi^*$ states for graphene are
degenerate at the K point corners of the BZ. This degeneracy occurs
at the Dirac point energy which coincides with the Fermi level at
half-filling, resulting in a point-like Fermi surface.
The DCA works very well for doping levels less
than $\approx300$ meV and allows one to describe the 2D plasmon
properties correctly in the fields of graphene plasmonics and
photonics with respect to experimental data
\cite{kochnl11,chn12,fen12,grponp12,frlonc13,gaacsp14,rolis15}.

\medskip
\par

However, in recent papers
\cite{stscprb10,gayussc11,denoprb13,pisinjp14}, where realistic
energy band dispersion of graphene was taken into account, some
anisotropy in the 2D plasmon dispersion at $T=0$ was reported for
doped graphene. Additionally, it was found
\cite{gayussc11,pisinjp14,pochc14} that for momentum transfer along
the $\Gamma$K direction, a distinctive second plasmon branch evolves
from the Dirac point at zero temperature as the Fermi level is
located at energies exceeding $\approx 300$ meV. Characteristically,
this mode starts to appear at high frequency and extends downward as
its intensity is increased at longer wavelength. Its origin was
traced to the trigonal warping of the  energy bands forming the
Dirac cone in the graphene band structure \cite{slwepr58,mcc69}.
This results in a two-component
 electron system with different Fermi velocities for which such an additional
mode may exist \cite{picjp56,sigaepl04}.

\medskip
\par

\medskip
\par

While a square-root-like dispersion of the 2D plasmon has been
reported  for free-standing graphene and graphene
adsorbed on dielectric substrates
\cite{chn12,fen12,liwiprb08,pflajpcm11,shhwprb11,shkiapl11,lapfnjp12},
experiments carried out on monolayer graphene grown on metals
exhibit a linear dispersion due to screening experienced by a
metallic substrate. Specifically, experiments for graphene on
Pt(111) \cite{pomaprb11,pomaprb12, pofojpcm13,pochn14} and on
Ir(111) \cite{lafonjp11} have shown a nearly identical dispersion
relation \cite{hoiu15}. The work of Ref. \cite{jaraprb11} concerns
evaluating the probability function for single layer graphene and
comparing it with the integrated (over wave vector) energy loss
spectra in Ref. \cite{gomcan13}.

\medskip
\par

Generally, the theoretical and experimental study of plasmon modes
of free-standing graphene and graphene-metals at finite temperature
merits special attention. This would be a key step toward
engineering plasmonic application of graphene. However, so far, the
theoretical investigation of the impact of temperature on the 2D
plasmon in graphene in the framework of a DCA was performed
\cite{daliprb13}. Whereas the high-energy $\pi$ and $\pi+\sigma$
plasmons are marginally affected by $T$, it was shown that
temperature plays an important role in the low-frequency dielectric
properties of graphene since the density of free carriers $\rho>0$
at $T\neq0$ due to a gapless energy spectrum. Due to a linear energy
dispersion of the $\pi$ and $\pi^*$ bands  in the vicinity of the
Dirac point, the 2D plasmon frequency at low $T$ goes
\cite{daliprb13} as $\omega_{2D}\sim T^{1/2}$ maintaining a
characteristic for a 2D system \cite{stprl67} $\sim q^{1/2}$
dependence on the momentum magnitude as well. At the same time,
there has been no detailed theoretical investigation of plasmons
involving the temperature and anisotropy of graphene considering its
electronic band structure beyond the DCA. In fact, electron
collective excitations, apart from their dependence on the magnitude
of transferred momentum, may also strongly depend on its direction,
and the doping could further affect the anisotropy. The results we
obtain in this paper demonstrate the anisotropy of the plasmon
spectrum along symmetry directions within the BZ and would be
suitable for verification in experiments where the transferred
momentum or temperature is held fixed.

\medskip
\par

Here, we  report on a theoretical
investigation of the anisotropy of low-energy graphene plasmon
excitations which may be induced by finite temperature in either the
presence or absence of carrier doping. Adjusting the chemical
potential with the use of an electric field effect, we observe an
unusual plasmon second branch whose intensity and linewidth at
finite temperature may differ from its zero temperature counterpart.
This paper is organized as
follows: In Sec.~II, we describe details of the {\it ab initio}
calculation of the graphene dielectric and loss functions. The
calculated results and their discussion are reported in Sec.~III.
The main conclusions of this work are given in Sec.~IV. Unless
otherwise stated explicitly, atomic units ($\hbar=e^2=m_e=1$) are
used throughout the paper.

\section{Calculation details}

Excitations in an electron system are characterized by the
transferred momentum ${\bf q}$ and excitation frequency $\omega$,
which determine the dielectric function \cite{pino66}. In this work
we calculate the dielectric function for free-standing graphene  at
arbitrary temperature $T$ for several choices of the chemical
potential $\mu$, including the $\mu=0$ case. Our starting point is
the electronic band structure evaluated in a periodically repeated
(and well separated) graphene sheets geometry. Based on such a
three-dimensional (3D) geometry the imaginary part of the density
response function of non-interacting electrons in reciprocal space
is expressed as

\begin{widetext}
\begin{equation}\label{spectral_function}
{\rm Im}[\chi^0_{{\bf G}{\bf G}'}({\bf q},\omega)] =
\frac{2}{\Omega} \sum^{\rm BZ}_{\bf k} \sum_{n n'} (f_{n{\bf
k}}-f_{n'{\bf k}+{\bf q}}) \langle n{\bf k}|e^{-{\rm i}({\bf q}+{\bf
G})\cdot{\bf r}}|n'{\bf k}+{\bf q}\rangle \langle n'{\bf k}+{\bf
q}|e^{{\rm i}({\bf q}+{\bf G}')\cdot{\bf r}}|n{\bf k}\rangle
\delta(\varepsilon_{n{\bf k}}-\varepsilon_{n'{\bf k}+{\bf
q}}+\omega).
\end{equation}
\end{widetext}
In this notation,  ${\bf G}\equiv\{{\bf g},g_z\}$ is a 3D reciprocal
lattice vector, ${\bf g}$ and ${\bf q}$ are in-plane 2D reciprocal
lattice vector and wave vector in the first BZ, respectively. In Eq.
(\ref{spectral_function}) the factor 2 accounts for spin, $\Omega$
is a normalization volume, the sum over wave vectors ${\bf k}$ is
performed in the first BZ, $n$ and $n'$ are the energy-band indices,
$f_{n{\bf k}}=1/\left[e^{(\varepsilon_{n{\bf k}}-\mu)/T}+1\right]$ are the
temperature-dependent Fermi occupation factors, and $\mu$ is the
chemical potential. The eigenvectors $|n {\bf k}\rangle$ and
eigenenergies $\varepsilon_{n{\bf k}}$ are the self-consistent
solution of the Kohn-Sham Hamiltonian of the density functional
theory taking the exchange-correlation potential in the
Ceperley-Alder form \cite{cealprl80}. The Troullier-Martin nonlocal
norm-conserving ionic potential \cite{trmaprb91} was taken for
description of the electron-ion interaction. In Eq.~(\ref{spectral_function}) we employed a 720$\times$720$\times$1
${\bf k}$ mesh for summation over the BZ. In numerical calculations
performed by using our own code \cite{sichprb03}, the
$\delta$-function in Eq.~(\ref{spectral_function}) was represented
by a Gaussian with a broadening parameter of 25 meV. The real part
of $\chi^0$ was obtained from Im$[\chi^0]$ via the Kramers-Kronig
relation by numerical integration. For this,  the Im$[\chi^0]$
matrices were calculated on a discrete energy mesh in the 0-20 eV
interval with a step of 1 meV.

\medskip
\par

In order to proceed, one should obtain the density response function
for interacting electrons $\chi$ which in the time-dependent density
functional theory \cite{rugrprl84,pegoprl96} obeys the integral
Dyson equation, $\chi=\chi^0+\chi (\upsilon+K_{\rm xc})\chi^0$,
where $\upsilon$ is the Coulomb potential and $K_{\rm xc}$ accounts
for the dynamical exchange-correlations. In this work we employed
the random-phase approximation (RPA) \cite{ehcopr59}, i.e. setting
$K_{\rm xc}$ to zero. A major problem in solving the Dyson equation
for the artificial 3D superlattice used here is the appearance of a
spurious long-range Coulomb interaction between the collective
charge oscillations in the different graphene sheets.
Whereas such interaction does not
influence notably the calculated properties of the conventional
surface and acoustic plasmons \cite{sichprl04,dipon07,yajaprb12}, it
alters severely the 2D plasmon dispersion which becomes
qualitatively wrong in the ${\bf q} \rightarrow 0$ limit
\cite{besiprb03}.  In order to solve this problem we followed the
recipe proposed recently by Nazarov \cite{nanjp15}. As a result of
such a procedure, in the evaluated $\chi_{{\bf G}{\bf G}'}({\bf
q},\omega)$ matrix the spurious Coulomb interaction between graphene
sheets is eliminated and the respective inverse dielectric function
defined as
\begin{equation}
\epsilon^{-1}_{{\bf G}{\bf G}'}({\bf q},\omega)=\delta_{{\bf G}{\bf
G}'}+\upsilon_{{\bf G}{\bf G}'}({\bf q}) \chi_{{\bf
G}{\bf G}'}({\bf q},\omega)
\end{equation}
contains information regarding the electron excitations in a single
free-standing graphene sheet only. For such $\epsilon^{-1}$, we
evaluated the 2D dielectric function $\epsilon({\bf q},\omega)$ and
the corresponding energy-loss function $L({\bf q},\omega)\equiv {\rm
Im}[\epsilon^{-1}({\bf q},\omega)]$ used in the following for
analysis of the excitation spectrum of graphene.
A peak in the energy loss function
may be identified as a collective mode or single-particle-like
excitation. The former corresponds to a zero point in
Re$[\epsilon(\textbf{q},\omega)]$ at which
Im$[\epsilon(\textbf{q},\omega)]$ is zero or very small, while the
latter relates to a finite value of
Im$[\epsilon(\textbf{q},\omega)]$. In particular, the plasmon energy
at certain ${\bf q}$ corresponds to the energy position of the
dominating sharp peak in the loss function $L({\bf q},\omega)$. The
peak is a function of the in-plane wave  vector ${\bf q}$ and
temperature. The critical momentum value $q_c$ is defined as that at
which the plasmon dispersion passes and enters the single-particle
mode excitation region, and is clearly identified at very  low
temperature. Also, it is useful to analyze the behavior of the
corresponding dielectric function $\epsilon({\bf q},\omega)$, since
the plasmon occurs at the energy at which the conditions
Re$[\epsilon]=0$ and Im$[\epsilon]=0$ (or presence of a local
minimum in Im$[\epsilon]$) are realized.

\medskip
\par

It is important to note that the chemical potential position depends on temperature.
At low $T$, this dependence is approximately given as
\begin{equation}
\mu \approx  E_F  \left[ 1  -  \frac{\pi^2}{6} \frac{d\ln(\rho(E_F ))}{d\ln(E_F )}  \left(
\frac{k_BT}{E_F }\right)^2 + \cdots  \right],
\end{equation}
where $\rho(E_F )$ is the density-of-states at the Fermi energy $E_F $.
For graphene, we  have $\rho(\varepsilon)=\varepsilon/[\pi(\hbar v_F)^2] $,
where $v_F$ is the Fermi velocity. Furthermore, in calculating the temperature-dependence
of the $\chi^0$ at low temperature ($k_BT\ll E_F $), one can apply
$f_0(\varepsilon;T)\approx  \theta(E_F -\varepsilon)- (k_BT)\delta(\varepsilon-E_F )$
in terms of the Heaviside step-function $\theta(x)$.

\medskip
\par

In evaluating $\chi^0$ at finite-temperature, the following transformation
of Maldague  \cite{LaryG2++} relating its values in the absence ($T=0$) and
presence of a heat bath can be employed, i.e.,

\begin{equation}
\chi^0(q,\omega;T) = \int_0^\infty d\,E
\frac{\chi^0_{T=0,E_F=E  }(q,\omega  )}{4 k_BT
\cosh ^2\left[ \frac{E-\mu(T)   }{2 k_B T}
\right]  }  \ .
\end{equation}

\medskip
\par

This approach is useful for carrying out analytical calculations at
low $T$ as was demonstrated in the case of graphene
\cite{daliprb13}. However, it does not provide any advantage in the
numerical calculations based on incorporation of the full energy
band structure since it requires evaluation of $\chi^0$, at least,
at several Fermi level positions. Therefore, in the current work we
numerically calculate the $\chi^0$ matrices according to Eq.~(\ref{spectral_function}) explicitly taking into account the finite
$T$ via the Fermi occupation factors. The extrinsic doping level
variation is simulated by placing the chemical potential in a given
position.

\section{Calculation Results}

\begin{figure*}
\includegraphics[width=1\textwidth,angle=0]{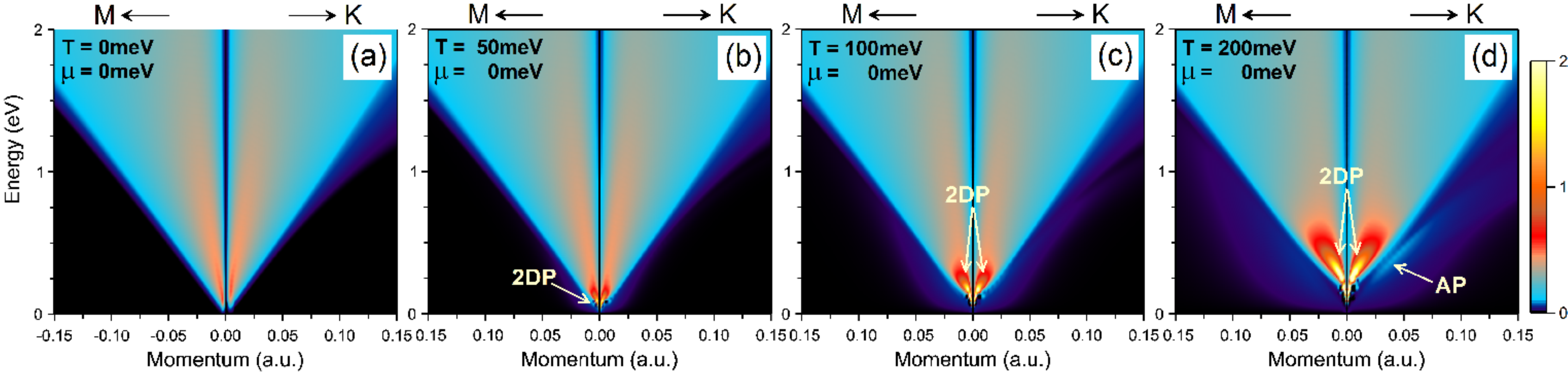}
\caption{ (Color online) Emergence of the 2D plasmon upon increasing
the temperature of the electron system in undoped graphene. The
excitation spectra, $L({\bf q},\omega)$, in the $\Gamma$M and
$\Gamma$K directions evaluated at temperatures of (a) 0, (b) 50, (c)
100, and (d) 200 meV are presented.
The borders of a region corresponding to the inter-band e-h pair
excitations are maintained almost intact upon variation of $T$,
whereas those for the intra-band e-h excitations rapidly expand with
$T$ increase.   A 2D plasmon (2DP) is observed at all chosen finite
temperatures with frequency dispersion $\omega_{2DP}\propto q^{1/2}$
at small $q$'s. } \label{fig_1}
\end{figure*}

In Fig.~\ref{fig_1}, we demonstrate how the 2D plasmon emerges in
undoped graphene when the temperature of its electron system is
increased. Figure \ref{fig_1}(a) shows that at $T=0$ the excitation
spectrum for low energies is governed by the incoherent e-h pairs
involving inter-band  transitions between the $\pi$ and $\pi^\ast$
energy bands which are the only excitations permitted in this energy
interval. However, at finite temperature the 2D plasmon starts to
emerge on the low-energy side. Thus, from Fig.~\ref{fig_1}(b),  it
is clear that at $T=50$ meV there is a peak in the loss function at
energies below approximately 150 meV. The dispersion of this mode
presents a clear behavior of a 2D plasmon mode, i.e. its energy goes
to zero with the momentum magnitude $q$ reduction as $\propto
q^{1/2}$ which is a characteristic of a 2D plasmon. However, there
is a substantial difference in comparison with the conventional 2D
plasmon case. It consists in that intrinsically the 2D plasmon peak
in the finite $T$ case has finite linewidth (e.g. finite lifetime)
at any $q$, whereas in the conventional 2D electron gas
\cite{stprl67} there is an energy threshold below which the 2D
plasmon has infinite lifetime (zero linewidth). This difference is
explained by the fact that in the finite-$T$ case the restrictions
on the phase-space for the e-h pair excitations are relaxed in
comparison with the $T=0$ case with the free-carriers
\cite{stprl67,anformp82}. From Fig.~\ref{fig_1}(b) it is evident
that the finite width of the 2D plasmon peak is due to decay into
inter-band e-h pairs
 since its dispersion occurs above the intra-band e-h pair
continuum, as it appears in a conventional 2D electron gas at
$T=0$.  However, even though at finite $T$ there is not a
well-defined low-energy border for the inter-band region, it is seen
in Fig.~\ref{fig_1}(b) how the 2D plasmon width is increased as its
energy is increased. Finally, at $\omega$ around 150 meV, the decay
into the inter-band e-h pairs becomes so efficient that at larger
energies this mode ceases to be a well-defined collective
excitation.

\medskip
\par

Upon increasing the temperature,  the number of free carriers in the
system is also increased.  This fact is reflected in the 2D
plasmon dispersion relation which is blue shifted as one can see from comparison
of the T=50, 100, and 200 meV plots presented in Figs.~\ref{fig_1}(b),
(c), and (d), respectively. Even though the region for the
intra-band e-h excitations is expanded when $T$ is increased, the
dispersion relation of 2D plasmons occurs over an expanded phase-space as $T$
grows. Therefore, in Figs.~\ref{fig_1}(c) and (d), one can resolve a well-defined
peak corresponding to the 2D plasmon up to energies of $\approx350$
and $\approx650$ meV for the T=100 and 200 meV cases, respectively.

\medskip
\par

From Fig. \ref{fig_1}, it is easy to perceive that the shape of the e-h continua
for the intra- and inter-band e-h pair excitations are visibly
anisotropic in this momentum-energy range. However, the 2D
plasmon dispersion relation for the corresponding temperatures is almost
isotropic and is in good agreement with results from a simple model employing the
DCA \cite{daliprb13}. This may be explained by noting that the 2D plasmon dispersion
relation for small momenta transfer is determined \cite{stprl67} by the total
number of free carriers which does not depend on the momentum direction for the
Dirac cone at low energies and the role played by other factors such as the
inter-band transitions is of less importance.

\medskip
\par

\begin{figure*}
\includegraphics[width=1\textwidth,angle=0]{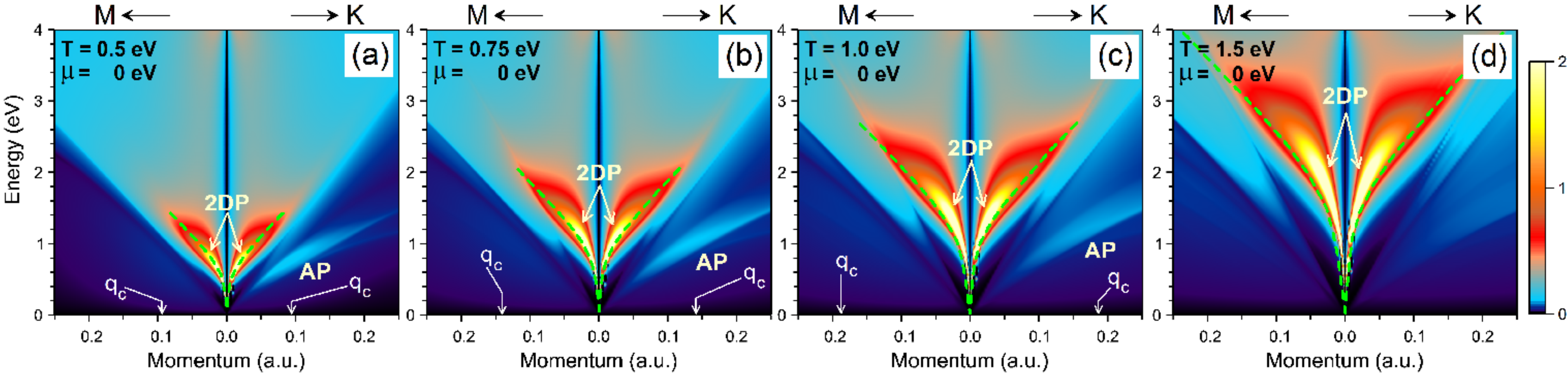}
\caption{ (Color online) Evolution of the 2D plasmon dispersion
relation upon increasing the temperature of the electron system in
undoped monolayer graphene. The excitation spectra,  $L({\bf
q},\omega)$, in the $\Gamma$M and $\Gamma$K directions evaluated at
$T$ equal to (a) 0.5, (b) 0.75, (c) 1.0, and (d) 1.5 eV. A 2D
plasmon (2DP) is observed for all these chosen temperatures with
energy dispersion $\omega_{2DP}\propto q^{1/2}$. Green  dashed lines
demonstrate the 2D plasmon dispersion obtained in the DCA
\cite{daliprb13} and occurring at $q$'s smaller than a cut-off
momentum $q_c$. } \label{fig_2}
\end{figure*}

Upon further $T$ increase, there are significant changes in the
excitation spectrum of undoped graphene. These are produced by the
strong trigonal warping of the energy bands forming the Dirac cone
at energies exceeding $\approx 300$ meV. In Fig.~\ref{fig_2},  we
compare the excitation spectrum calculated at $T=0.5$, 0.75, 1.0,
and 1.5 eV. By noting the dissimilarity between  Fig.~\ref{fig_2}(a)
with Fig.~\ref{fig_1}(d), one notices that an increase in $T$ from
200 meV to 500 meV drastically increases the size of the region for
the intra-band e-h transitions with a corresponding increase in the
number of free carriers in the systems. This results in an increase
in the 2D plasmon strength at $T=0.5$ eV accompanied by a notable
upward shift in its energy. Despite the increase in the number of
e-h excitations in this region, the 2D plasmon peak is well defined
over an extended energy range up to $\approx 1.5$ eV. In Fig.~\ref{fig_2}, we also plot the 2D dispersion evaluated in the
framework of a DCA. One can see in Fig.~\ref{fig_2}(a) that our 2D
plasmon dispersion relation almost coincides with
the DCA model predictions of Ref.~\cite{daliprb13}. Moreover, the momentum range where this mode can
be a well-defined collective excitation is almost the same in both
models. This demonstrates that the DCA is capable of describing the
2D plasmon behaviors even at such elevated temperatures. However,
upon further increase of $T$, notable deviations in our calculated
2D plasmon dispersion relations from the simple model predictions
become noticeable. Thus,  in Fig.~\ref{fig_2}(b), one can see that our calculated 2D plasmon
dispersion relation goes to slightly higher energy in comparison
with the DCA model curve represented by green dashed lines.
Moreover, at such elevated temperatures, some anisotropy in the 2D
plasmon dispersion becomes evident. Upon further temperature
increase, these main deviations of our 2D plasmon dispersion
relation  from the simple model curves increase as evidenced from
the loss function for $T=1.0$ and 1.5 eV reported in Figs.~\ref{fig_2}(c) and \ref{fig_2}(d), respectively. Thus, in the
$T=1.5$ eV case of Fig. \ref{fig_2}(d), our calculated 2D energy
dispersion exceeds that obtained in the simple model by about 15
$\%$. We attribute this deviation of our 2D plasmon dispersion to
the predictions in Ref.~\cite{daliprb13} to the presence of other
energy bands in the graphene band structure which are thermally
populated at such elevated $T$'s. As a result, the number of
carriers at finite $T$ is increased additionally due to a partial
occupation of the $\sigma$ bands, high density-of-states in the
$\pi$ and $\pi^*$ bands in the vicinity of the M point, and image
states \cite{sizhprb09}. Also, the anisotropy of the 2D plasmon
dispersion becomes more notable with the temperature increase.  This
behavior can also be explained by increased deviation of the density
of states in graphene \cite{pisinjp14} at increasing energy
separation from the Dirac point in comparison with that assumed in
the DCA.

\medskip
\par

Close examination of Fig.~\ref{fig_1}(d) reveals that for momentum
transfers in a 0.02-0.07 a.u. range along the $\Gamma$K direction a
faint peak referred to as an acoustic plasmon (AP) mode appears at
energies below the 2D plasmon. We explain its presence by the fact
that at such temperature some amount of slow carriers can be excited in
addition to those moving with $v_F \approx0.4$ a.u. as can be seen in Fig.~4(d)
of Ref.~\cite{pisinjp14}. As a result, a two-component electron gas
scenario may be realized in graphene at this $T$ even in the case
when the chemical potential is pinned at the Dirac point. As
demonstrated by the models, the dispersion of the AP mode
\cite{picjp56} should have an acoustic-like law, i.e., its energy
decays as $\sim q$ upon reduction of momentum magnitude $q$. From
Fig.~\ref{fig_1}(d), it is clear that the AP plasmon dispersion
relation in graphene
 deviates significantly from such behavior. Nevertheless, in order to stress its origin we apply the term
``acoustic plasmon" to this mode as well. When the temperature is increased, the number
of both slow and fast carriers increases as well, making the phenomenon of the AP
even more clear as evidenced from the loss spectrum for $T=0.5$ eV of
Fig.~\ref{fig_2}(a), where an AP peak can be discerned along the $\Gamma$K at $q$'s
exceeding $\approx0.03$ a.u.

\begin{figure}
\includegraphics[width=0.45\textwidth,angle=0]{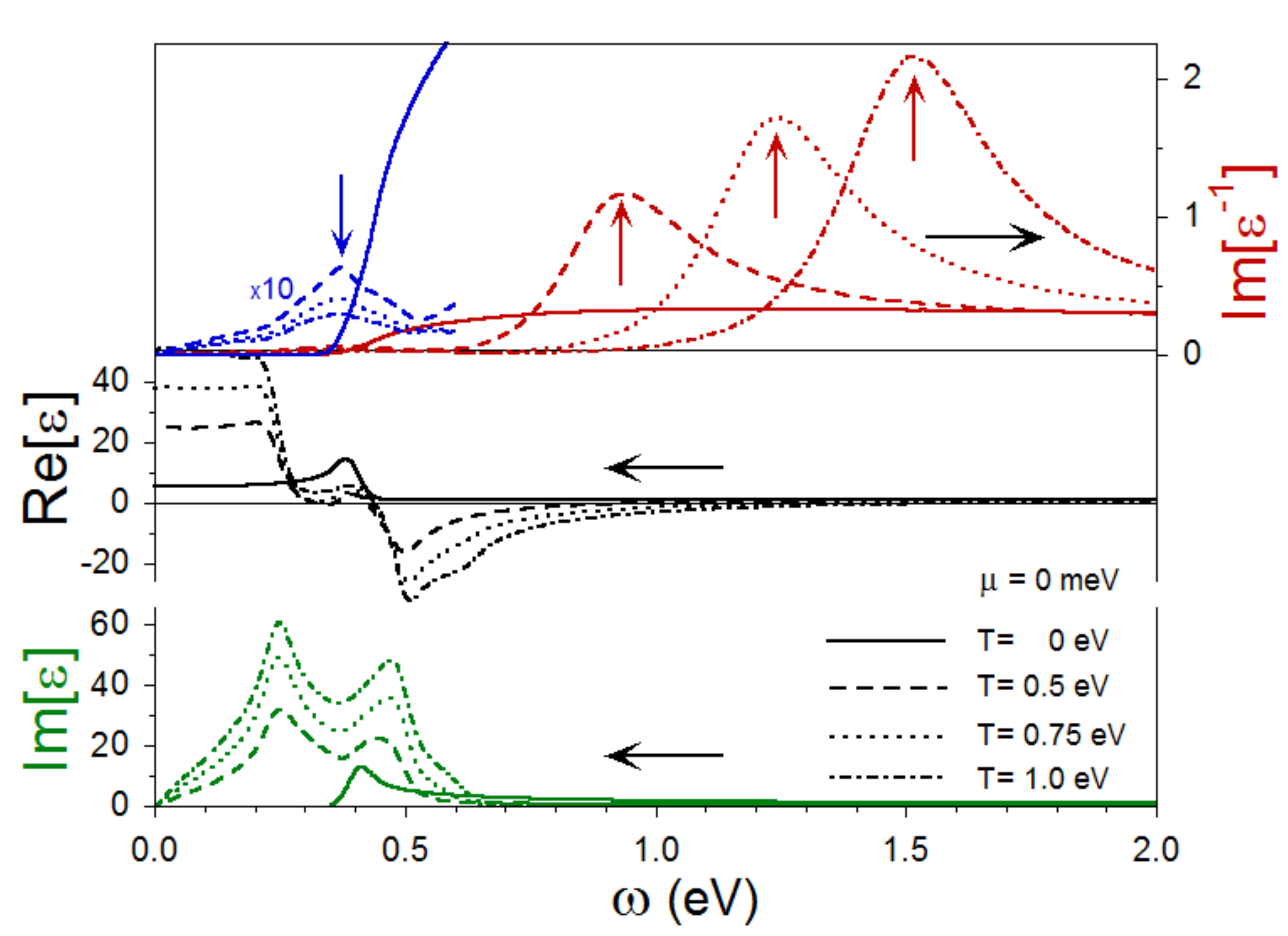}
\caption{ (Color online) The real and imaginary parts of the
dielectric function $\epsilon({\bf q},\omega)$ as well as the energy
loss function $L({\bf q},\omega)$ for some values of temperature at
$\mu=0$. The data are for wave vector directed along the $\Gamma$K
directions with $q=0.038$ a.u.   In
the upper panel at energies below 0.6 eV the blue lines show the
loss function multiplied by a factor 10. Blue vertical arrow marks
the energy position of the AP mode. Red vertical arrows mark the
energy positions of the 2DP.}  \label{fig_B1}
\end{figure}

\medskip
\par

In order to demonstrate the origin of the AP mode,
we present  in Fig.~\ref{fig_B1}
the $\omega$-dependence of the real and imaginary parts of the
dielectric function as well as the loss function calculated for
$\mu=0$ and various temperatures $T$ at $q=0.038$ a.u. along
$\Gamma$K. The zeros of Re$[\epsilon]$ correspond to the plasmon
excitation frequencies while the Im$[\epsilon]$ gives the Landau
damping.  As a matter of fact, an undamped plasmon mode occurs at a
frequency when ``{\em both}" the real and imaginary parts of the
dielectric function are zero. However at the chosen temperatures the
imaginary part of the dielectric function is zero only for $T=0$ at
energies below the threshold $\omega_t=0.35$ eV for the interband
$\pi$-$\pi^*$ transitions for this ${\bf q}$. In all other cases, it
is finite at the non-zero $\omega$'s. Nevertheless, the energy loss
function displays peaks, each arising at a frequency where there is
either a 2D plasmon or an AP mode. The height of each peak for the
loss function is representative of the intensity of the plasmon
excitation mode. We note that the imaginary part of $\epsilon({\bf
q},\omega)$  in Fig. \ref{fig_B1}
becomes larger on the low-energy side as the temperature is
increased, which reflects the increasing number of charge carriers.
There is a corresponding modification of the real part. A
consequence of this behavior is a rapid increase in the energy of
the 2D plasmon when the temperature is raised. In the low-energy
part of Im$[\epsilon]$ at $T=0.5$ eV of Fig.~\ref{fig_B1} one can
see that instead of a single peak
seen at $\omega=0.42$ eV in the $T=0$ case there are two peaks at
energies below 1 eV. Their appearance is related, as explained above,
to the intra-band transitions involving  two kinds of carriers
moving with different group velocities in the $\Gamma$K direction
\cite{pisinjp14}. This makes the real part of $\epsilon$ cross the
zero axis additionally twice in this energy interval. Such
zero-crossings together with existence of a local minimum in
Im$[\epsilon]$ leads to the appearance of a peak in the loss function at
$\omega=0.35$ eV that signals the existence of the AP mode at this
energy. Upon the increase in temperature, the strength of both peaks
in Im$[\epsilon]$ gradually increases due to an increased number of
free carriers in the system. This is accompanied by the upward shift
of Re$[\epsilon]$ at $\omega$ less than 0.45 eV as seen in the
curves corresponding to $T=0.75$ and 1.0 eV. Consequently, the
condition for the existence of a well-defined collective
AP mode is relaxed. This is
confirmed by a gradual reduction of the spectral weight of the AP
peak in the loss function at these $T$'s.

\medskip
\par

\begin{figure}
\includegraphics[width=0.45\textwidth,angle=0]{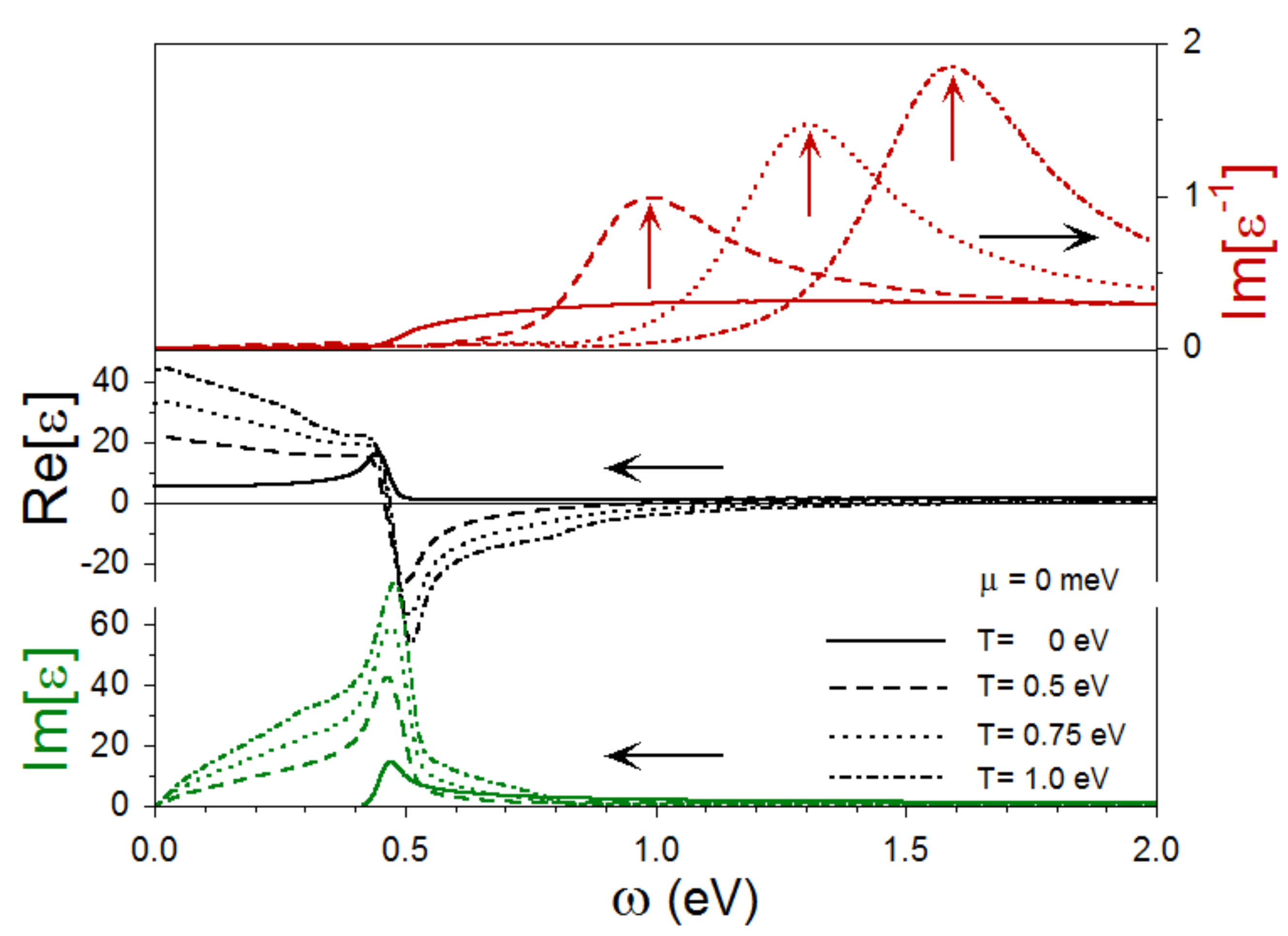}
\caption{ (Color online) The real and imaginary parts of the
dielectric function $\epsilon({\bf q},\omega)$ as well as the energy
loss function $L({\bf q},\omega)$ for some values of temperature at
$\mu=0$. The data are for wave vector directed along the $\Gamma$M
directions with $q=0.044$ a.u. The red
vertical arrows mark the energy positions of the 2DP.}
\label{fig_B2}
\end{figure}

Viewing closely  the loss spectra  in Figs.~\ref{fig_1} and
\ref{fig_2} reveals that a peak corresponding to the AP mode does
not appear along the $\Gamma$M direction. As in the case for
extrinsic doping which was considered in Ref. \cite{pisinjp14} and
in what  follows, its absence in this direction is attributed to the
presence of only one kind of carriers moving in this direction for
the energy interval from -1 eV to +1 eV as seen in Fig.~4(d) of Ref.~\cite{pisinjp14}. This is confirmed by Fig.~\ref{fig_B2} where we
report the real and imaginary parts of the dielectric function and
the loss function calculated for $\mu=0$ and various temperatures
$T$ at $q=0.044$ a.u. along $\Gamma$M. Since, in this direction
there is only one kind of carriers, the Im$[\epsilon]$ presents only
a single intra-band peak at finite $T$'s. Consequently, the real
part of $\epsilon$ crosses the zero axis only once at energies
around $\omega=0.45$ eV where Im$[\epsilon]$ possesses a peak. As a
result, the corresponding mode cannot be realized since it is
Landau-damped. This is confirmed by the absence of any peak in the
nearby energy region in the loss function. Instead,
in it only the peak corresponding
to the conventional 2DP mode is observed at higher energies.

\begin{figure*}
\includegraphics[width=1\textwidth,angle=0]{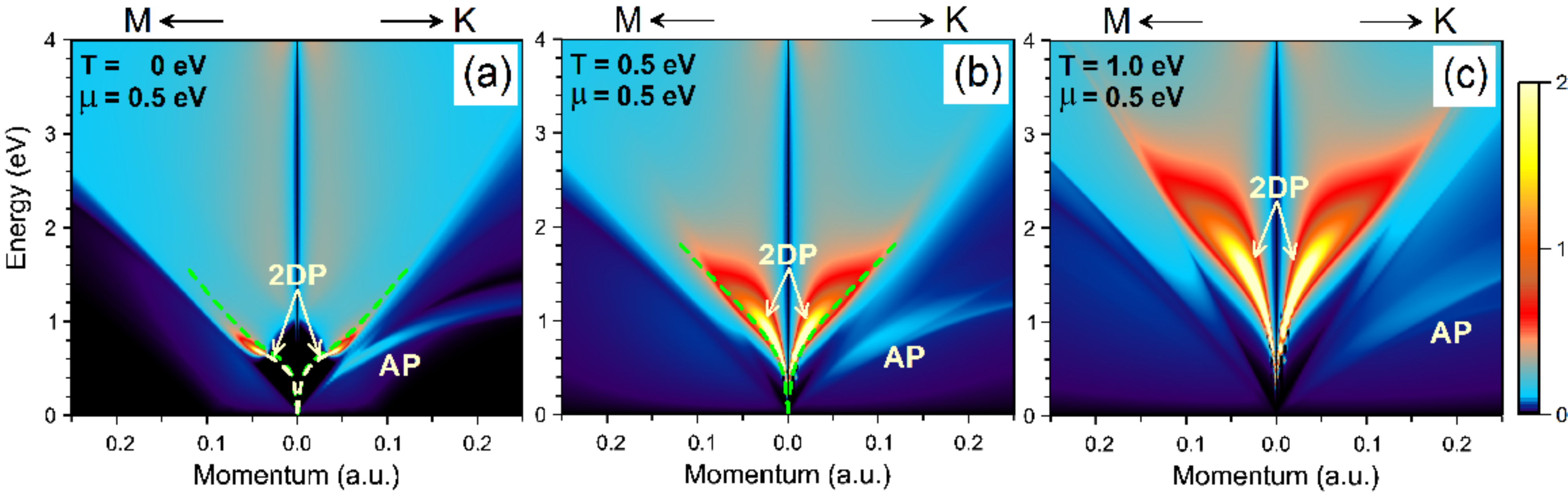}
\caption{ (Color online) Excitation spectra,  $L({\bf q},\omega)$,
in the $\Gamma$M and $\Gamma$K directions evaluated at $\mu=0.5$ eV
and $T$ of (a) 0, (b) 0.5 and (c) 1.0 eV. The 2D plasmon (2DP) and
acoustic plasmon (AP) dispersions are denoted by corresponding
labels. Yellow dashed lines show
the undamped 2DP dispersion. Green dashed lines demonstrate the 2D
plasmon dispersion obtained in the DCA model \cite{daliprb13}. }
\label{fig_3}
\end{figure*}

\begin{figure*}
\includegraphics[width=1\textwidth,angle=0]{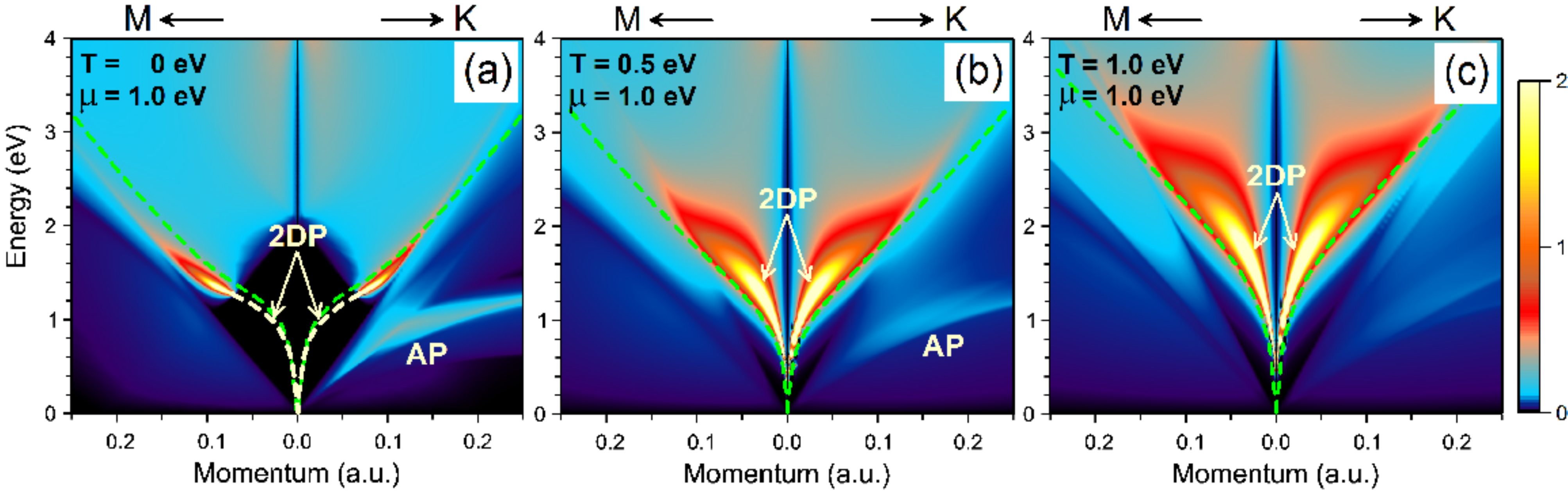}
\caption{ (Color online) Excitation spectra,  $L({\bf q},\omega)$,
in the $\Gamma$M and $\Gamma$K directions evaluated at $\mu=1.0$ eV
and $T$ of (a) 0, (b) 0.5 and (c) 1.0 eV. The 2D plasmon (2DP) and
acoustic plasmon (AP) dispersions are denoted by corresponding
labels. Yellow dashed lines show
the undamped 2DP dispersion. Green dashed lines demonstrate the 2D
plasmon dispersion obtained in the DCA model \cite{daliprb13}.}
\label{fig_4}
\end{figure*}

\begin{figure*}
\includegraphics[width=1\textwidth,angle=0]{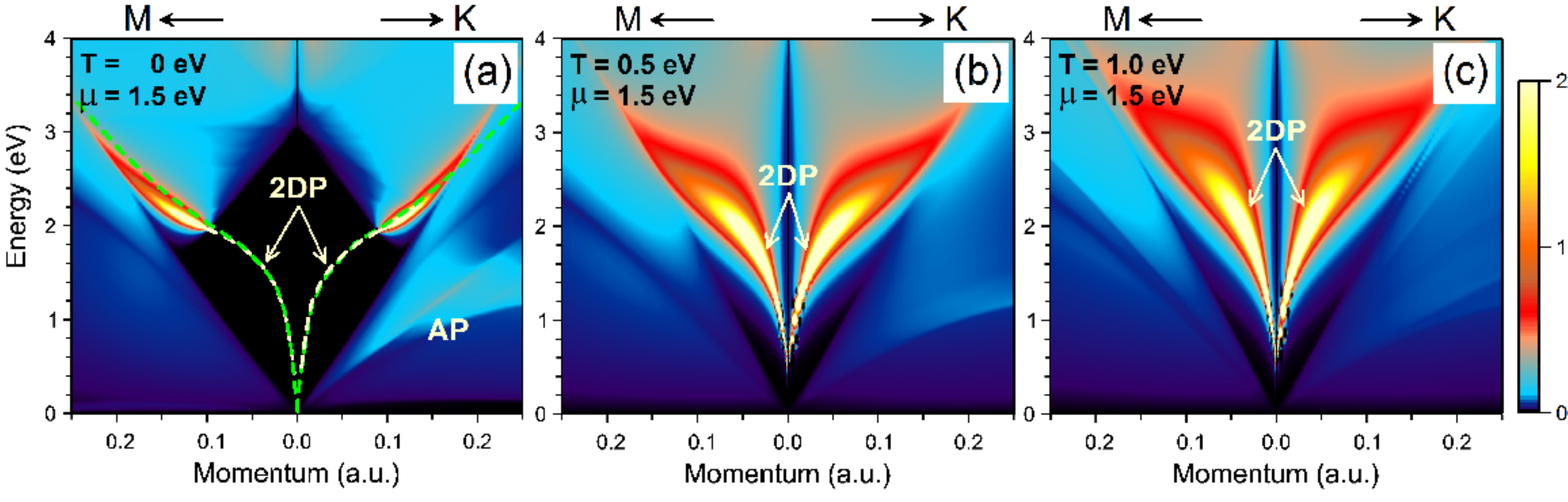}
\caption{ (Color online) Excitation spectra,  $L({\bf q},\omega)$,
in the $\Gamma$M and $\Gamma$K directions evaluated at $\mu=1.5$ eV
and $T$ of (a) 0, (b) 0.5  and (c) 1.0 eV. The 2D plasmon (2DP) and
acoustic plasmon (AP) dispersions are denoted by corresponding
labels.  Yellow dashed lines show
the undamped 2DP dispersion. Green dashed lines demonstrate the 2D
plasmon dispersion obtained in the DCA model \cite{daliprb13}.}
\label{fig_5}
\end{figure*}

\begin{figure}
\includegraphics[width=0.45\textwidth,angle=0]{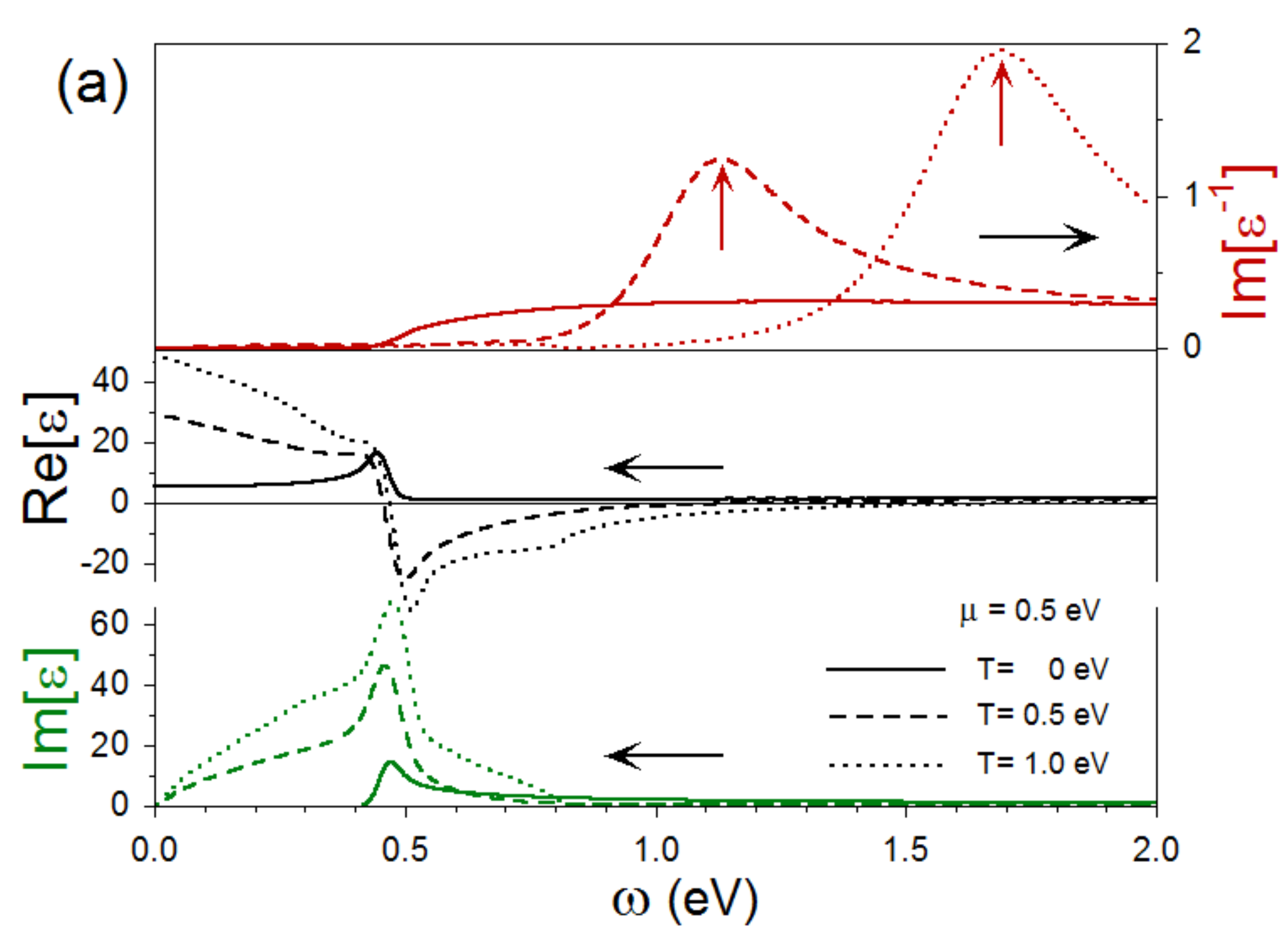}\\
\includegraphics[width=0.45\textwidth,angle=0]{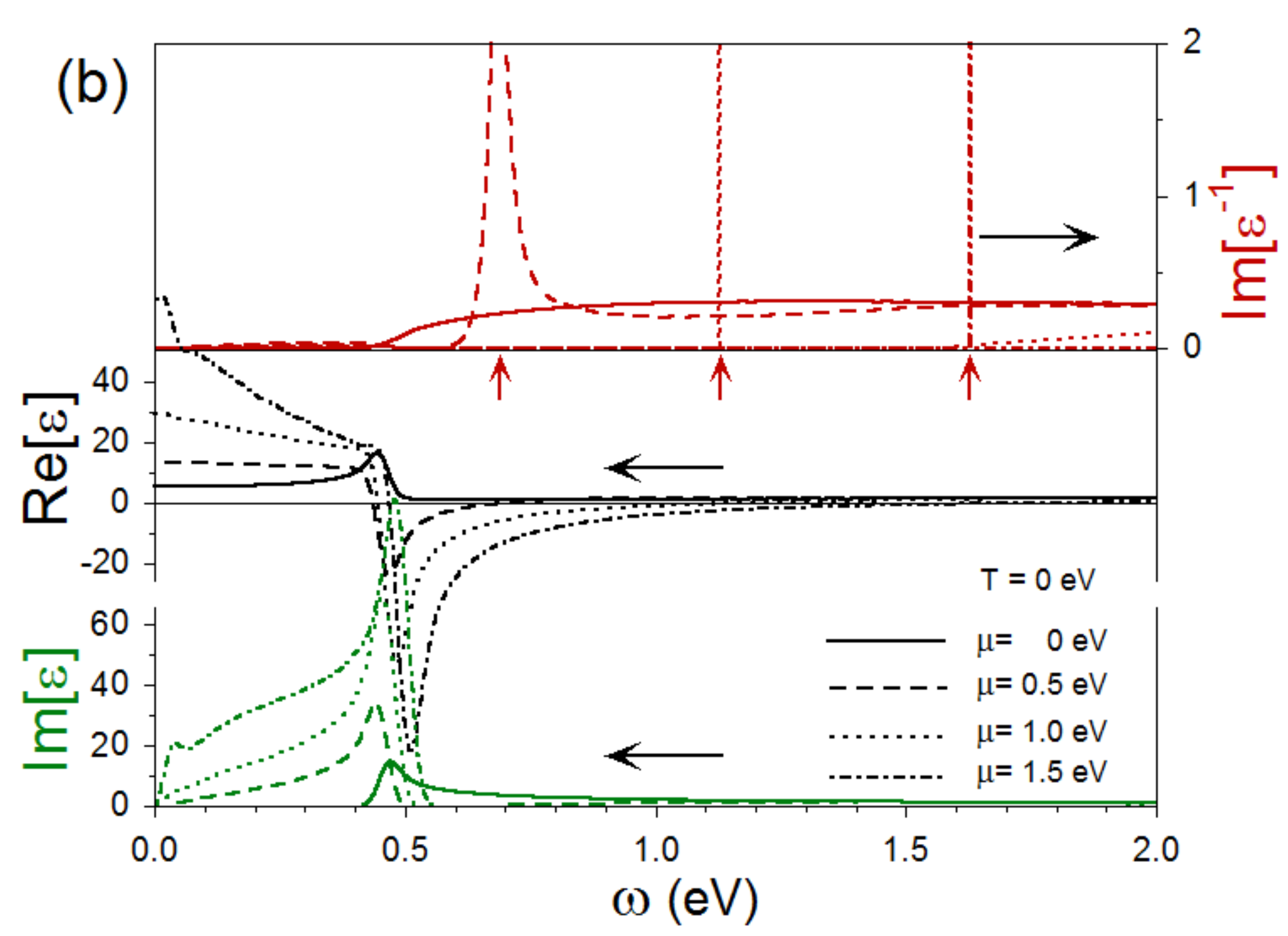}\\
\includegraphics[width=0.45\textwidth,angle=0]{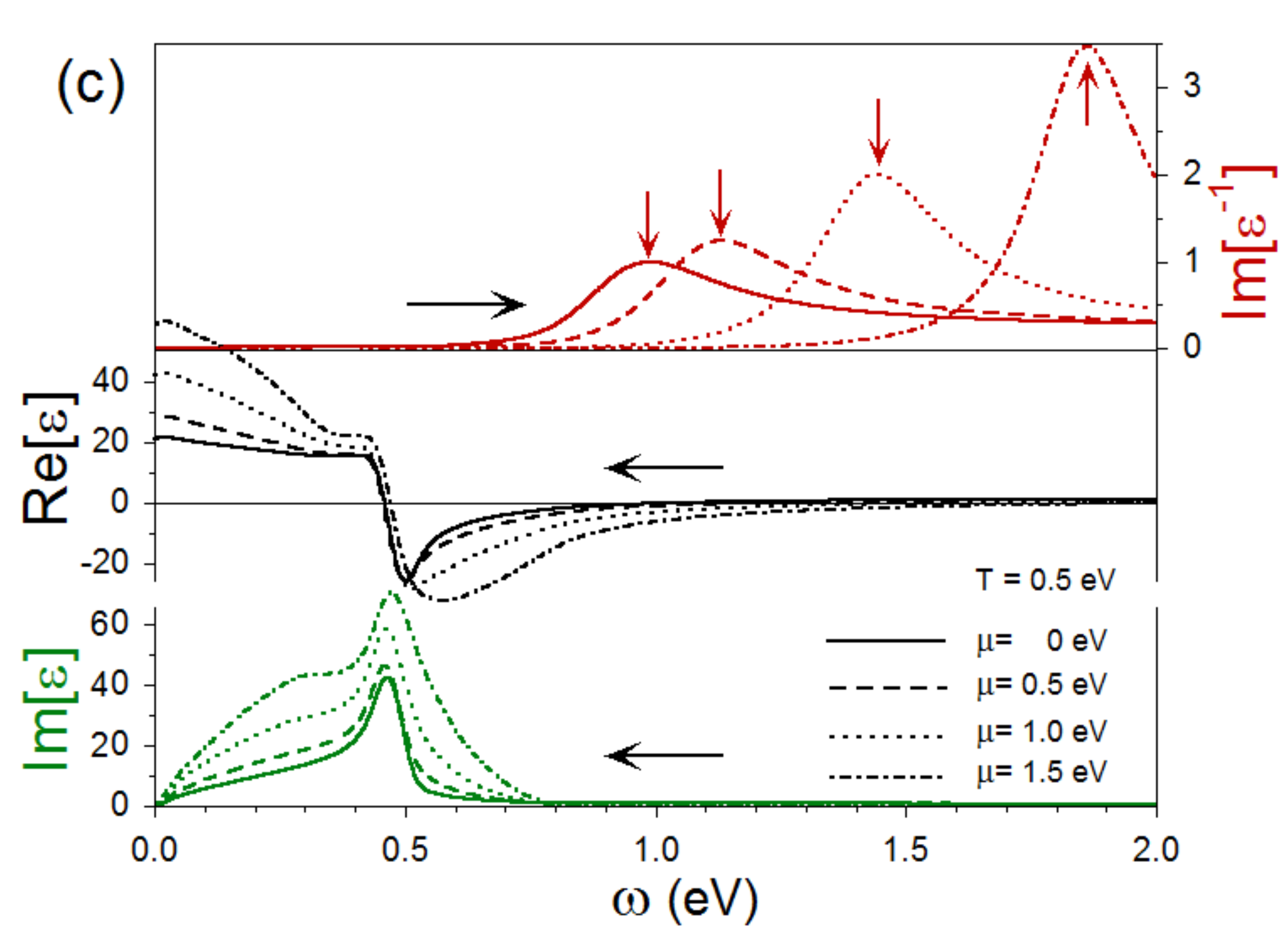}
\caption{ (Color online) The real and imaginary parts of the
dielectric function $\epsilon({\bf q},\omega)$ as well as the energy
loss function,  $L({\bf q},\omega)$, for some values of doping and
temperature. The data are for wave vector directed along the
$\Gamma$M direction with $q=0.044$. The chemical potential level
$\mu$ and temperature $T$ are as follows: (a) $\mu=0.5$ eV and
various $T$, (b) $T=0$ and various $\mu$, (c) $T=0.5$ eV and various
$\mu$. Red vertical arrows mark the
energy positions of the 2DP.}
\label{fig_6}
\end{figure}

\medskip
\par

It is interesting to compare the spectra reported in Fig.~\ref{fig_2} with those corresponding to the cases with non-zero
doping $\mu$ upon the temperature change. These data for $\mu=0.5$,
1.0, and 1.5 eV are reported in Figs.~\ref{fig_3}, \ref{fig_4} and
\ref{fig_5}, respectively. Figures  \ref{fig_3}(a), \ref{fig_4}(a),
and \ref{fig_5}(a) demonstrate the behavior of the energy-loss
spectra at zero temperature as the doping level is increased from
$0.5$ to $1.5$ eV, i.e., expanding the $0-1.0$ eV energy range
explored at zero temperature in previous publication
\cite{pisinjp14}. From the $1.5$ eV panel in Fig.\ \ref{fig_5}(a),
it is clear that the 2D plasmon also exists even at such a high
doping level. At small momenta, the 2D plasmon dispersion frequency
is enhanced due to an increased carriers number at the corresponding
Fermi level. In contrast to the loss spectra for $\mu=0$ and
$T\neq0$ of Fig.~\ref{fig_2} the 2D plasmon peak in Figs.
\ref{fig_3}(a), \ref{fig_4}(a), and \ref{fig_5}(a) has zero width
over the extended phase space momentum-energy region, as highlighted
by yellow dashed line. One can observe how this region where the 2D
mode is undamped is gradually increased upon shifting the position
of $\mu$ at energy as high as 1.5 eV. At such doping level the 2D
plasmon does not decay into e-h pairs up to energy of $\approx2$ eV.
In Figs.~\ref{fig_3}(a),
\ref{fig_4}(a) and \ref{fig_5}(a) with green dashed lines, we show the
2DP dispersion predicted by the DCA model \cite{daliprb13} which clearly
does not depend on the momentum direction. From comparison
of the 2DP dispersion calculated here with these curves one can
observe its notable asymmetry. Moreover, apparently the momentum
(and energy) range where this mode is a well defined collective
excitation is significantly smaller than that predicted employing
the DCA. It may signal that this model underestimates the decay rate
of the 2DP into e-h pairs.

\medskip
\par

Additionally, from the excitation
spectrum reported in Fig.~\ref{fig_5}(a), one can deduce that the AP
mode also exists at $\mu=1.5$ eV along the $\Gamma$K direction, even
though the width of the corresponding peak in the loss function is
notably increased in comparison to the $\mu=0.5$ and 1.0 eV cases
indicating that with the doping level greater than $1.0$ eV, this
mode more efficiently decays into incoherent e-h pairs. Its
dispersion is linear at values of $q$ reaching $0.25$ a.u. which is
in contrast to the $\mu=0.5$ and 1.0 eV cases, where it starts to
deviate from the linear dispersion at significantly lower $q$'s.

\medskip
\par

Figures \ref{fig_3}, \ref{fig_4}, and \ref{fig_5}    explore  the ``{\em combined\/}"  effect of temperature and doping
on the Coulomb excitations in monolayer graphene. These results clearly  demonstrate the novelty and
importance of our studies through    our  modeling and extensive numerical calculations. Our comprehensive studies
 predict an unusual and diverse  behavior of the many-body properties for doped and undoped graphene at
finite temperature. There is a 2DP mode in all panels for both
$\Gamma$M  and  $\Gamma$K directions for all values of doping and
temperature studied here.    For some
finite values of $T$ our calculated 2DP dispersion can be compared
with the one predicted by the DCA approximation of \cite{daliprb13}.
In general, the agreement between two models is rather good,
although a prominent anisotropy in the 2DP dispersion obtained in
the full calculations is evident. Another observation in Figs.~\ref{fig_3}, \ref{fig_4} and \ref{fig_5} consists in that at finite
$T$'s the 2DP dispersion obtained here is always blue shifted in
comparison that predicted by the DCA. We explain such behavior by
the deviation of graphene band structure from the DCA upon energy
separation from the Dirac point. Interestingly, in the $T=0$ case of
Figs.~\ref{fig_3}, \ref{fig_4}, and \ref{fig_5}, our 2DP dispersion
over extended momentum  range goes at slightly lower energies in
comparison with that generated by the DCA model.

\medskip
\par

In Fig.~\ref{fig_6}(a), we present some detailed results for the
real and imaginary parts of the dielectric function along with the
energy loss function for doping $\mu=0.5$ eV and various
temperatures. Again, the location of the peak in the loss function
for the 2D plasmon excitation energy gets shifted upward rapidly as
the temperature is increased due to a substantial increase in the
Drude peak of Im$[\epsilon]$,  which is indicative of a swift
increase in the number of carriers that are thermally excited. This
may be accounted for by the corresponding  shift in the
density-of-states away from the Dirac point. The dependence of the
dielectric properties of graphene upon doping at $T=0$ K is also
worth investigating. Figure \ref{fig_6}(b) (again, data for
$q=0.044$ a.u. along $\Gamma$M) shows this for undoped graphene
along with  four chosen doping values. Regarding the combined effect
arising from temperature and doping on the 2D plasmon excitation
spectrum, we present Fig.~\ref{fig_6}(c). In general,  when $T$ and
$\mu$ are increased, the 2D plasmon frequency is increased.

\medskip
\par

In Figs. \ref{fig_3}, \ref{fig_4}, and \ref{fig_5} one can observe
how the AP peak in the loss function corresponding to the acoustic
plasmon mode gradually washed out from the calculated loss spectra
upon the $T$ increase. In Figs.~\ref{fig_3}, \ref{fig_4}, and
\ref{fig_5}, the AP branch is suppressed when the values for the
doping and temperature are simultaneously large. Additionally, one
can see that the intensity of the AP branch is decreased as  the
doping level exceeds 1.0 eV. A general tendency is that this mode
ceases to exist at a lower $T$ value as the doping level is
increased. Thus, in Fig.~\ref{fig_3} at $\mu=0.5$ eV the AP peak can
be detected at all the temperature reaching 1.0 eV. In Fig.
\ref{fig_4} one can see that at $\mu=1.0$ eV the AP peak is well
defined at $T=0$, being strongly damped at $T=0.5$ eV. Its presence
in the $T=1.0$ eV spectrum is barely visible, demonstrating that can
not exist at such combination of $\mu$ and $T$. From Fig.~\ref{fig_5} corresponding to case of the 1.5 eV doping, the AP peak
is completely removed from the $T=1.0$ eV loss spectrum of Fig.~\ref{fig_5}(c). At such elevated doping level the peak originally
corresponding to the AP mode is so broad at $T=0.5$ eV of Fig.~\ref{fig_5}(b) that it can not be considered as a well-defined
collective excitation. Disappearance of the AP mode in graphene at
significantly lower temperature upon increase of the doping level
(accompanied by the strong increase of number of carriers at the
Fermi level) correlates with effective destruction of the AP mode in
bulk Pd at significantly lower temperatures as was reported recently
\cite{sinaprb16}.

 \medskip
\par

\section{Concluding Remarks}
\label{sec5}

In summary, we have employed the  density functional method to
calculate the electronic energy bands for graphene in the absence of
an external  magnetic field. Using these results, we calculated  the
longitudinal wave vector and frequency-dependent dielectric function
at arbitrary temperature.  In our calculations, the entire energy
band  spectrum was included. This ensures the correctness of the
dielectric function  in the RPA and consequently the plasmon
intensity  and its frequency. We  presented the plasmon excitation
spectra at  various temperatures  and doping concentrations  which
may   be achieved experimentally. Our method may  also be applicable
to other rare 2D materials with Dirac cones such as silicene,
germanene and graphyne over a wide range of temperature
and doping. Our numerical results
may be validated by inelastic light-scattering or high-resolution
electron-energy loss spectroscopy which has been successfully
applied to the two-dimensional electron gas system.

\section{Acknowledgments}

V.M.S acknowledges the partial support from the  University of the
Basque Country UPV/EHU, Grant No. IT-756-13 and
the Spanish Ministry of Economy and Competitiveness MINECO, Grant
No. FIS2016-76617-P.


\begin{thebibliography}{99}


\bibitem{drprb74} G. Dresselhaus, Phys. Rev. B {\bf 10}, 3602
(1974).

\bibitem{najpsj76} K. Nakao, J. Phys. Soc. Jpn {\bf 40}, 761 (1976).

\bibitem{chmiprb91} J. C. Charlier, J. P. Michenaud, X. Gonze, and
J.  P. Vigneron, Phys. Rev. B {\bf 44}, 13237 (1991).

\bibitem{chgoprb91} J. C. Charlier,  X. Gonze, and J. P. Michenaud,  Phys. Rev. B {\bf 43}, 4579 (1991).

\bibitem{chmiprb92} J. C. Charlier, J. P. Michenaud, and X. Gonze, Phys. Rev. B {\bf 46}, 4531 (1992).

\bibitem{bupssb77} U. B\"uchner, Phys. Status Solidi (b) {\bf 81}, 227 (1977).

\bibitem{krhaprl08} C. Kramberger, R. Hambach, C. Giorgetti, M. H. Rummeli, M. Knupfer, J. Fink, B. B\"uchner, Lucia Reining, E. Einarsson, S. Maruyama, F. Sottile, K. Hannewald, V. Olevano, A. G. Marinopoulos, and T. Pichler, Phys. Rev. Lett. {\bf 100}, 196803 (2008).

\bibitem{luloprb09} J. Lu, K. P. Loh, H. Huang, W. Chen, and A. T.
S. Wee, Phys. Rev. B {\bf 80}, 113410 (2009).

\bibitem{kreiprb10} C. Kramberger, E. Einarsson, S. Huotari, T. Thurakitseree, S. Maruyama, M. Knupfer, and T. Pichler, Phys. Rev. B {\bf
81}, 205410 (2010).

\bibitem{buhoan13} S. Z. Butler {\it at al.}, ACS Nano {\bf 7},
2898 (2013).

\bibitem{xulicr13} M. Xu, T. Liang, M. Shi, and H. Chen, Chem. Rev.
{\bf 113}, 3766 (2013).

\bibitem{wapr47} P. R. Wallace, Phys. Rev. {\bf 71}, 622 (1947).

\bibitem{wustnjp06} B. Wunsch, T. Stauber, F. Sols, and F. Guinea, New J. Phys. {\bf 8}, 318 (2006).

\bibitem{hwdaprb07} E. H. Hwang and S. Das Sarma, Phys. Rev. B {\bf 75}, 205418 (2007).

\bibitem {kochnl11} F. H. Koppens, D. E. Chang, and F. J. Garcia de Abajo, Nano Lett. {\bf 11}, 3370 (2011).

\bibitem{chn12} J. Chen {\it el al.}, Nature {\bf 487}, 77 (2012).

\bibitem{fen12} Z. Fei {\it el al.}, Nature {\bf 487}, 82 (2012).

\bibitem{grponp12} A. N. Grigorenko, M. Polini, and K. S. Novoselov, Nature Phot. {\bf 6}, 749 (2012).

\bibitem{frlonc13} M. Freitag, T. Low, W. J. Zhu, H. G. Yan, F. N.
Xia, and P. Avouris, Nature Commun. {\bf 4}, 1951 (2013).

\bibitem{gaacsp14} F. J. Garcia de Abajo, ACS Phot. {\bf 1}, 135 (2014).

\bibitem{rolis15} D. Rodrigo, O. Limaj, D. Janner, D. Etezadi, F. J.
Garcia de Abajo, V. Pruneri, and H. Altug, Science {\bf 349}, 165
(2015).

\bibitem{stscprb10} T. Stauber, J. Schliemann, and N. M. R. Peres,
Phys. Rev. B {\bf 81}, 085409 (2010).

\bibitem{gayussc11} Y. Gao and Z. Yuan, Solid State Commun. {\bf 151}, 1009 (2011).

\bibitem{denoprb13} V. Despoja, D. Novko, K. Dekani\ifmmode \acute{c}\else \'{c}\fi{}, M.  \ifmmode \check{S}\else \v{S}\fi{}unji\ifmmode \acute{c}\else \'{c}\fi{},  and L. Maru\ifmmode \check{s}\else \v{s}\fi{}i\ifmmode \acute{c}\else \'{c}\fi{}, Phys. Rev. B {\bf 87}, 075447 (2013).

\bibitem{pisinjp14} M. Pisarra, A. Sindona, P. Riccardi, V. M. Silkin, and J. M. Pitarke, New J. Phys. {\bf 16}, 083003 (2014).

\bibitem{pochc14} A. Politano and G. Chiarello, Carbon {\bf 71}, 176
(2014).

\bibitem{slwepr58} J. C. Slonczewski and P. R. Weiss, Phys. Rev. {\bf
109}, 272 (1958).

\bibitem{mcc69} J. W. McClure, Carbon {\bf 7}, 425 (1969).

\bibitem{picjp56} D. Pines, Can. J. Phys. {\bf 34}, 1379 (1956).

\bibitem{sigaepl04} V. M. Silkin, A. Garcia-Lekue, J. M. Pitarke, E. V. Chulkov, E. Zaremba,
and P. M. Echenique, Europhys. Lett. {\bf 66}, 260 (2004).

\bibitem{liwiprb08} Y. Liu, R. F. Willis, K. V. Emtsev, and T.
Seyller, Phys. Rev. B {\bf 78}, 201403 (2008).

\bibitem{pflajpcm11} H. Pfn\''{u}r, T. Langer, J. Baringhaus, and C. Tegenkamp, J. Phys.: Condens. Matter {\bf 23}, 112204 (2011).

\bibitem{shhwprb11} S. Y. Shin, C. G. Hwang, S. J. Sung, N. D. Kim,
H. S. Kim, and J. W. Chung, Phys. Rev. B {\bf 83}, 161403(R) (2011).

\bibitem{shkiapl11} S. Y. Shin, N. D. Kim, H. S. Kim, K. S. Kim, D. Y. Noh, K. S. Kim,
 and J. W. Chung, Appl. Phys. Lett. B {\bf 99}, 082110 (2011).

\bibitem{lapfnjp12} T. Langer, H. Pfn\"ur, C. Tegenkamp, S. Forti, K. Emtsov, and U. Starke, New J. Phys. {\bf 14}, 103045 (2012).

\bibitem{pomaprb11} A. Politano, A. R. Marino, V. Formoso, D. Fari\'as, R. Miranda, and
G. Chiarello, Phys. Rev. B {\bf 84}, 033401 (2011).

\bibitem{pomaprb12} A. Politano, A. R. Marino, and G. Chiarello, Phys. Rev. B {\bf 86}, 085420 (2012).

\bibitem{pofojpcm13} A. Politano, V. Formoso, and G. Chiarello, J. Phys.: Condens. Mat. {\bf 25}, 345303 (2013).

\bibitem{pochn14} A. Politano and G. Chiarello, Nanoscale {\bf 6}, 10927 (2014).


\bibitem{lafonjp11} T. Langer, D. F. Forster, C. Busse, T. Michely, and H. Pfn\"ur, New J. Phys. {\bf 13}, 053006 (2011).

\bibitem{hoiu15} N. J. M. Horin, A. Iurov, G. Gumbs, A. Politano, and G. Chiarello,
{\it Recent Progress in Nonlocal Graphene Plasmons in
Low-Dimensional and Nanostructured Materials and Devices} (Springer,
2015) pages 205-237.

\bibitem{jaraprb11} V. Borka Jovanovi\'c, I. Radovi\'c, D. Borka, and Z. L. Mi\v{s}kovi\'c, Phys. Rev. B {\bf 84}, 155416 (2011).

\bibitem{gomcan13} C. Gong, S. McDonnell, X. Qin, A. Azcatl, H. Dong, Y. J. Chabal, K. Chao,
and R. M. Wallace, ACS Nano {\bf 8}, 642649 (2013).

\bibitem{daliprb13} S. Das Sarma and Q. Li, Phys. Rev. B {\bf 87}, 235418 (2013).

\bibitem{stprl67} F. Stern, Phys. Rev. Lett. {\bf 18}, 546 (1967).

\bibitem{pino66} D. Pines and P. Nozi\`eres, {\it The Theory of Quantum Liquids: Normal Fermi Liquids} (W. A. Benjamin, New York, 1966), Vol.
1.

\bibitem{cealprl80} D. M. Ceperley and B. J. Alder, Phys. Rev. Lett. {\bf 45}, 566
(1980), as parametrized by J. P. Perdew and A. Zunger, Phys. Rev. B
{\bf 23}, 5048 (1981).

\bibitem{trmaprb91} N. Troullier and J. L. Martins, Phys. Rev. B {\bf 43}, 1993 (1991).

\bibitem{sichprb03} V. M. Silkin, E. V. Chulkov, and P. M. Echenique, Phys. Rev. B {\bf 68}, 205106 (2003).

\bibitem{rugrprl84} E. Runge and E. K. U. Gross, Phys. Rev. Lett. {\bf 52}, 997 (1984).

\bibitem{pegoprl96} M. Petersilka, U. J. Gossmann, and E. K. U. Gross, Phys. Rev. Lett. {\bf 76}, 1212 (1996).

\bibitem{ehcopr59} H. Ehrenreich and M. H. Cohen, Phys. Rev. {\bf 115}, 786 (1959).

\bibitem{sichprl04} V. M. Silkin, E. V. Chulkov, and P. M. Echenique, Phys. Rev. Lett. {\bf 93}, 176801 (2004).

\bibitem{dipon07} B. Diaconescu, K. Pohl, L. Vattuone, L. Savio, P. Hofmann, V. M. Silkin,
J. M. Pitarke, E. V. Chulkov, P. M. Echenique, and M. Rocca, Nature
{\bf 448}, 57 (2007).

\bibitem{yajaprb12} J. Yan, K. W. Jacobsen, and K. S. Thygesen, Phys. Rev. B {\bf 86}, 241404(R) (2012).

\bibitem{besiprb03} A. Bergara, V. M. Silkin, E. V. Chulkov, and P. M. Echenique, Phys. Rev. B {\bf 67}, 245402 (2003).

\bibitem{nanjp15} V. U. Nazarov, New J. Phys. {\bf 17}, 073018 (2015).

\bibitem{LaryG2++}  P. F. Maldague, Surf. Sci. {\bf 73}, 296 (1978).

\bibitem{anformp82} T. Ando, A. B. Fowler, and F. Stern, Rev. Mod. Phys. {\bf 54}, 437 (1982).

\bibitem{sizhprb09} V. M. Silkin, J. Zhao, F. Guinea, E. V. Chulkov, P. M. Echenique, and H. Petek, Phys. Rev. B {\bf 80}, 121408 (2009).

\bibitem{sinaprb16} V. M. Silkin, V. U. Nazarov, A. Balassis, I. P. Chernov, and E. V. Chulkov, Phys. Rev. B {\bf 94}, 165122 (2016).

\end{thebibliography}
\end{document}